\documentclass[12pt,letterpaper,fleqn]{article}
\usepackage[latin1]{inputenc}
\usepackage{amsmath}
\usepackage{amsfonts}
\usepackage{amssymb}
\usepackage{graphicx}
\usepackage[left=2cm,right=2cm,top=2cm,bottom=2cm]{geometry}
\author{ \renewcommand{\thefootnote}{\alph{footnote}}
Jason Gilbert\footnotemark[1] \quad and \quad Cameron Baribeau \vspace{0.5cm}\\
\small \emph{Canadian Light Source, University of Saskatchewan, Saskatoon, Canada}\vspace{0.2cm}
}
\title{Measurement of magnetic fields using the voltage generated by a vibrating wire}

\newcommand{\ps}{\ + \ }
\newcommand{\ms}{\ - \ }
\newcommand{\es}{\ = \ }
\newcommand{\bsin}[1]{\sin\left( #1 \right)}	
\newcommand{\bcos}[1]{\cos\left( #1 \right)}
\newcommand{\fulld}{\mathrm{d}}

\usepackage{subcaption}
\usepackage{float}

\usepackage{siunitx}
\DeclareSIUnit\gauss{G}

\begin{document}
\renewcommand{\thefootnote}{\alph{footnote}}
\footnotetext[1]{Corresponding author. Electronic mail: jtg074@usask.ca}
\maketitle
\numberwithin{equation}{section}
\numberwithin{figure}{section}

\begin{abstract}
A vibrating wire may be used as an instrument with a variety of applications, one of which is the measurement of magnetic fields. Often, the magnetic fields are determined by measuring the amplitude of the wire vibration under the action of a Lorentz force. Though generally adequate, this approach may be inconvenient in certain circumstances. One of these occurs when it is necessary to measure the amplitude of high-frequency vibration, as the amplitude is expected to decrease linearly with frequency, and thus becomes harder to measure. Another example may be found in situations where the sensor must operate over a wide range of vibration frequencies. In this case the sensor will be unresponsive to specific frequencies of wire vibration, which are determined by the placement of the sensor. This means that for the instrument to be robust, the sensor must be precisely mobile, or multiple sensors must be used. 

Here a technique which may be used to supplement the displacement sensor is described. This technique makes use of the voltage generated by the motion of the wire in the magnetic field under measurement. It is predicted that the technique may be more suitable for measurements requiring high frequency vibration, and is sensitive to all frequencies of vibration. Measurements of a magnetic field obtained using this technique are compared to those found using only a displacement sensor, and the benefits and drawbacks of the technique are discussed.
\end{abstract}

\newpage
\section{Introduction}
In the field of accelerator physics the vibrating wire technique \cite{temnykh1997vibrating} is utilized to measure the magnetic field of a variety of devices. These include quadrupoles \cite{temnykh1999magnetic, wolf2005vibrating, wouters2012wire, vrankovic2014wire}, undulators \cite{temnykh2010study}, and wigglers \cite{temnykh2001some, temnykh2001chess, temnykh2003vibratingCESR, temnykh2003vibrating, baribeau2019magnetic}.

This technique has recently been used at the Canadian Light Source \cite{baribeau2019magnetic} in an attempt to measure the magnetic field along the centerline of an insertion device, a hybrid in-vacuum wiggler with a period of \SI{80}{\milli\metre}. While taking these measurements, it was noted that the electrical impedance of the wire appeared to be changing. It was proposed that this change may be explained by an electromotive force (emf) generated by the motion of the wire. To test the validity of this explanation, the hypothetical emf generated by the wire was calculated, and in doing so, it was found that it may be possible to obtain useful information about the magnetic field from measurements of the emf. For an example of this principle applied to the measurement of fluid viscosity, see Ref. \cite{padua1998electromechanical}.

This contribution is organized as follows. In Section \ref{sec:Theory} the theory of the vibrating wire technique is briefly explained; after which, a description of the emf generated by the vibration of a wire, in terms of this theory, is developed. In Section \ref{sec:Application} the application of this emf for magnetic measurement is discussed. In Section \ref{sec:Measurement and Analysis} methods used to test the practicality of this approach, and the results of these tests, are discussed. Finally, the conclusions drawn from this work are summarized in Section \ref{sec:Conclusion}.

\section{Theory} \label{sec:Theory}
\subsection{A summary of the vibrating wire technique}
To provide context for the measurement method being proposed, the conventional method must first be described. This description will be minimal, and those who are interested in more detail may refer to Ref. \cite{temnykh1997vibrating}.

The principle of the vibrating wire technique is that by measuring the displacement of a wire due to a Lorentz force acting along its length, the magnetic field which causes the Lorentz force may be inferred. It is well known that the transverse displacement of a taut wire under the influence of some driving force is described by
\begin{equation} \label{eq:diffEq}
Tu_{zz} \ms \gamma u_t \ps \mu u_{tt} \es F \ .
\end{equation}
Here $u = u(t, z)$ is the transverse displacement of the wire, and $F = F(t, z)$ is the force per unit length acting on the wire; $t$ is the time at which the wire is observed, and $z$ is the direction which the wire extends. Subscripts are used to denote differentiation with respect to the subscripted symbol. The coefficients of the differential terms are $T$ the tension in the wire, $\mu$ the mass per unit length of the wire, and $\gamma$ the damping coefficient.

\begin{figure}[H]
\centering
\includegraphics[width=0.50\textwidth]{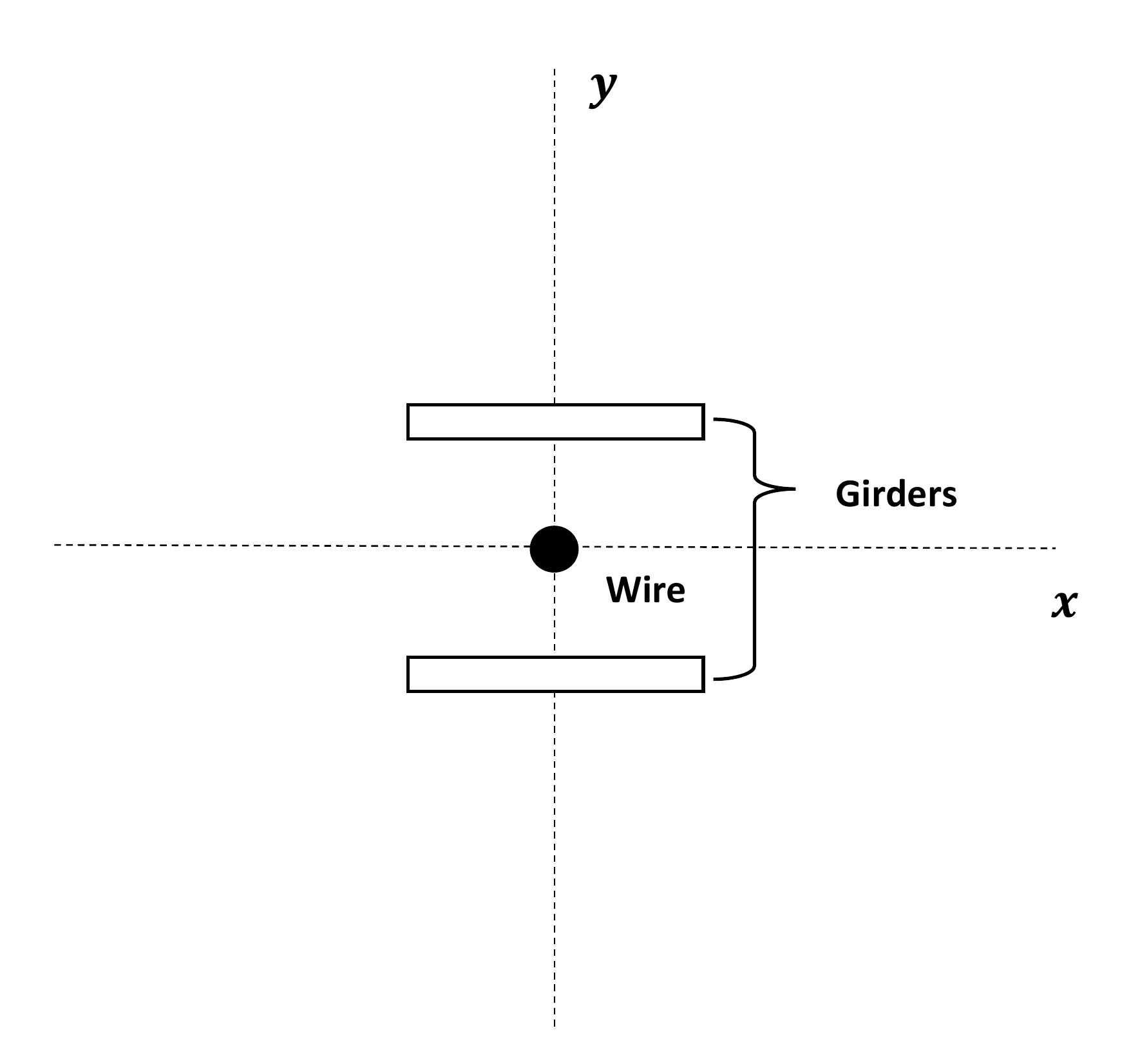}
\caption{A cross section of the measurement apparatus. The wire extends in the $z$ direction, perpendicular to the page. The magnetic field produced by the girders is in the $y-$direction. The motion of the wire is ideally confined to the plane $y=0$.}
\label{fig:ProblemSchematic}
\end{figure}

In the context of magnetic field measurement, this differential equation relates a Lorentz force $F(t, z)$ acting on the wire to the transverse displacement $u(t, z)$ of the wire; if the current in the wire is known, the magnetic field acting on the wire may be determined by measuring the displacement of the wire. A diagram which defines the coordinate system, depicting an insertion device, may be found in Figure \ref{fig:ProblemSchematic}.

It has been shown \cite{temnykh1997vibrating} that if the wire is held fixed at points which are outside the influence of the magnetic field, a solution to the above differential equation may be obtained in the form of a sine-series. This means that the magnetic field may be expressed as 
\begin{equation} \label{eq:serCoeff}
B(z) \es \sum_{n=0}^{\infty} B_n \bsin{\dfrac{n\pi}{L}z} \ ,
\end{equation}
where the series coefficients $B_n$ are related to the displacement of the wire, and $L$ is the distance between the points where the wire is held fixed. If the wire is underdamped, and the current in the wire is of the form 
\begin{equation}
I(t) = I_0e^{i\omega t} \ ,
\end{equation}
which is meant to represent a sinusoidal signal with amplitude $I_0$ and angular frequency $\omega$, written in the form of a complex exponential for convenient mathematical manipulation, then the series coefficients are related to the vibration of the wire by
\begin{equation} \label{eq:wireDisp}
u_n(t, z) \es B_n \bsin{\dfrac{n\pi}{L} z} \dfrac{I_0 \bcos{\omega t - \phi}}{\mu\left[ (\gamma'\omega)^2 - (\omega^2 - \omega_n^2)^2 \right]^{1/2}} \ .
\end{equation}
Here $u_n$ is the displacement of the wire when driven at a frequency near the $n^{th}$ harmonic of the fundamental frequency of the wire; $\omega_n$ is the frequency of the $n^{th}$ harmonic, which corresponds to the $n^{th}$ series coefficient, and is given by
\begin{equation} \label{eq:resonantFrequency}
\omega_n \es \dfrac{n\pi}{L}\sqrt{\dfrac{T}{\mu}} \ ;
\end{equation}
$\gamma' = \gamma\mu^{-1}$, and $\phi$ is the phase shift between the driving current and the wire vibration, defined as
\begin{equation} \label{eq:phaseTem}
\tan\phi \es \frac{\gamma'\omega}{\omega_n^2 - \omega^2} \ .
\end{equation}

Examining equation \eqref{eq:wireDisp}, it can be seen that the measured amplitude of the wire vibration is given by the time-independent component of the equation,
\begin{equation} \label{eq:ampTem}
A_n \es B_n \bsin{\dfrac{n\pi}{L}z_s} \dfrac{I_0}{\mu}\left[ (\gamma'\omega)^2 - (\omega^2 - \omega_n^2)^2 \right]^{-1/2} \ ,
\end{equation}
where $z_s$ is the position of the displacement sensor. From this the series coefficients may be obtained by fitting the following equation to measurements of vibration amplitude,
\begin{equation} \label{eq:fitTem}
A_n \es a_n\left[ c_n\omega^2 - ( \omega^2 - b_n )^2 \right]^{-1/2} \ ,
\end{equation}
assuming that $T$, $\mu$, and $\gamma$ are effectively constant for the curve. From which the series coefficients may be found to be
\begin{equation} \label{eq:coeffTem}
|B_n| \es a_n \dfrac{\mu}{I_0}\bsin{\dfrac{n\pi}{L}z_s}^{-1} \ .
\end{equation}
One should note that equation \eqref{eq:ampTem}, from which the series coefficients are calculated, is strictly of one sign, as determined by the sensor position $z_s$. This means that it can not by itself be used to calculate the corresponding magnetic field via equation \eqref{eq:serCoeff}, as it must be known which of the coefficients are positive and which are negative. This information may be obtained from measurements of the phase shift between the wire vibration and the driving current, as was done in \cite{temnykh1997vibrating}.

To briefly elaborate on the need for assuming the wire is underdamped in the derivation of equation \eqref{eq:wireDisp}, it is required so that the curves of equation \eqref{eq:wireDisp} are sufficiently localized, so that each $u_n$ is near zero in the vicinity of $u_{n \pm 1}$. If this is not the case, equations \eqref{eq:ampTem}-\eqref{eq:coeffTem} will be better represented as a sum over $n$, and a more complicated approach to fitting may be required.

\subsection{Electromotive force as a signal} \label{sec:EMF}
In the previous section it was explained how an electrical signal may be used to induce motion in a wire by means of a Lorentz force, and that this motion acts as a signal which carries information about the strength of the magnetic field along the length of the wire. Here an alternative approach will be explored. In this case, the motion of the wire will be considered to be a way of generating an electrical signal, and this electrical signal will be shown to carry information about the magnetic field.

The fact that an electrical signal is generated by the vibration of the wire may be explained by Faraday's Law. The circuit containing the wire is a loop of conducting material, meaning the vibration of the wire causes the area of, and therefore the magnetic flux through, the circuit to vary over time, giving rise to a voltage via $\varepsilon = -\fulld \Phi / \fulld t$. This means that a voltage which opposes the driving current will be generated across the wire, so that the wire appears to resist the driving current. In terms of an electric motor, the signal we mean to discuss may be referred to as "counter-electromotive force" or "back emf". Here it will simply be referred to as the emf.

At this point a mathematical description of $\varepsilon$, the emf generated by a vibrating wire, will be derived for a wire located at the center of an insertion device. In the case of a wire positioned at $x=y=0$, and a magnetic field directed along the vertical direction $y$, such as is depicted in Figure \ref{fig:ProblemSchematic}, the magnetic field acting on the wire may be written $B_y(x, z) = B(x, 0, z)$, where the subscript indicates the component of the magnetic field. If the motion of the wire is confined to the plane $y=0$, then the emf can be expressed in terms of the flux enclosed by the circuit as
\begin{equation}
\varepsilon \es -\dfrac{\fulld \Phi}{\fulld t} \es -\dfrac{\fulld}{\fulld t}\int_0^{L}\int_{-\infty}^{u}B_y(x, z)\ \fulld x \fulld z \ ,
\end{equation}
where $u$ is the displacement of the wire. Since $u = u(t)$, by the fundamental theorem of calculus
\begin{equation} \label{eq:generalEmf}
\varepsilon \es -\int_0^{L} \dfrac{\partial u}{\partial t} B_y(u, z)\ \fulld z \ .
\end{equation}
If the displacement of the wire takes the form of a standing wave this becomes
\begin{equation} \label{eq:generalIntegral}
\varepsilon \es -\dfrac{\partial \Lambda}{\partial t} \int_0^{L} \bsin{\dfrac{n\pi}{L}z} B_y(u, z)\ \fulld z \ ,
\end{equation}
where $\Lambda = \Lambda(t)$ is the amplitude of the wire displacement at time $t$. Noting that $u = u(z)$, meaning $B_y(u, z) = B_y(z)$, and recalling the definition of the sine-series, 
\begin{equation}
B(z) \es \sum_{n=0}^{\infty}B_n \bsin{\dfrac{n\pi}{L}z}
\qquad
\texttt{and}
\qquad
B_n \es \dfrac{2}{L} \int_0^{L}B(z)\bsin{\dfrac{n\pi}{L}z}\ \fulld z \ ,
\end{equation}
then it becomes clear that
\begin{equation}
\varepsilon \es -B_n \dfrac{L}{2} \dfrac{\partial \Lambda}{\partial t} \ .
\end{equation}

At this point, one should recall that these results are obtained for a driving force in the form of $F=F(z)$, meaning that it is assumed that the magnetic field does not vary appreciably in either of the transverse directions within the span of the wire vibration. 

If the motion of the wire is described by equation \eqref{eq:wireDisp}, then $u_n = \Lambda\bsin{\frac{n\pi}{L}z}$, and
\begin{equation} \label{eq:emfTime}
\varepsilon(t, \omega) \es B_n^2 \dfrac{I_0L}{2\mu} \bsin{\omega t - \phi} \omega\left[ (\gamma'\omega)^2 - \left(\omega^2 - \omega_n^2 \right)^2 \right]^{-1/2} \ ;
\end{equation}

comparing the time-dependent factors of equations \eqref{eq:emfTime} and \eqref{eq:wireDisp}, it can be seen that the phase difference between the emf and the driving current will be $\phi_\varepsilon = \phi + \pi / 2$, so that
\begin{equation} \label{eq:emfPhase}
\tan\phi_\varepsilon \es \dfrac{\omega_n^{2} - \omega^2}{\gamma'\omega} \ .
\end{equation}

The series coefficient from equation \eqref{eq:serCoeff} may then be obtained using a curve fitting approach similar to that described in the previous section, by measuring the amplitude of the emf as opposed to the amplitude of the wire vibration. Applied to equation \eqref{eq:emfTime}, this gives
\begin{equation} \label{eq:emfFit}
\varepsilon_n(\omega) \es a_n \omega\left[ c_n\omega^2 - (\omega^2 - b_n )^2 \right]^{-1/2} \ ,
\end{equation}
and
\begin{equation} \label{eq:emfCoeff}
|B_n| \es \sqrt{a_n\dfrac{2\mu}{I_0L}} \ .
\end{equation}
These results suggest that measurements of the voltage across the wire may be used to obtain information about the magnetic field acting on the wire, in a manner similar that described in the previous section. 

There is, however, one key difference between the two which one must take note. This is the fact that measurement of the voltage can only be used to determine the magnitude of the series coefficients. In mathematical terms, this is evidenced by the relationship $\varepsilon_n \propto B_n^2$, which is independent of the sign of $B_n$. In physical terms, one may consider the fact the voltage generated by the wire depends only on the speed of the wire, meaning it is independent of the direction of the wire moves with respect to the driving current.

\section{Application} \label{sec:Application}
In this section the application of the theory described in Section \ref{sec:EMF} will be discussed. 

To begin, the shortcomings of using emf as a signal will be considered. The first is that it may be unwieldy as a measurand for non-planar magnetic fields. Though the derivation of equation \eqref{eq:emfCoeff} explicitly treats only the case of planar motion in a field of one component, equation \eqref{eq:generalEmf} should be valid for any wire displacement in any magnetic field if $B_y(x, z)$ is replaced with $B(x, y, z)$. It must be remarked that in the case that there is a significant field in both of the directions transverse to the motion of the wire, each component of the field will contribute to the voltage which is produced. This would mean that the emf would carry information regarding both field components, and special care would need to be taken to interpret the meaning of the signal. If optical displacement sensors are used, the contribution of each field component may be easier to determine, as the motion in each direction may be measured independently.

Next, it should be reiterated that measurement of the amplitude of the emf is not by itself sufficient to calculate the magnetic field distribution. Unlike those results, in this case the signal being measured is proportional to the square of the series coefficient. This means that the sign of the coefficient may not be determined from measurement of the emf alone, making it unsuitable for determining the field distribution over space via equation \eqref{eq:serCoeff}. Despite this, applications which do not require the sign of the coefficients still exits. One such is the determination of the magnitude of coefficients related to specific frequencies of vibration, which may be used to characterize error fields such as in Ref. \cite{temnykh2003vibrating}.

Turning now to the advantages of measuring the emf, the first is that it may be used to complement some of the shortcomings of a displacement sensor. For example, the sensor may be positioned near the node of one of the harmonics, meaning that it would be unable to detect the vibration of the wire at that frequency. To elaborate on the problematic nature of the sensor response, consider equation \eqref{eq:coeffTem} in the form 
\begin{equation}
B_n \es a_n\dfrac{\mu}{I_0}\bsin{\dfrac{n\pi}{L}z_s}^{-1} \ \propto \ \bsin{x}^{-1} \ ,
\end{equation}
and apply simple calculus-based error analysis to obtain 
\begin{equation}
\dfrac{\partial B_n}{\partial x} \ \propto \ B_n\cot{(x)} \ .
\end{equation}
This shows that for $(x \mod \pi) \ll 1$, or $n z_s / L$ near an integer, the uncertainty in the series coefficient value becomes very large. This uncertainty becomes significant when many series coefficients are needed, and the distribution of these coefficients is difficult to predict, such as in the detailed measurement of insertion device fields \cite{baribeau2019magnetic}. To compensate for this, the sensor would need to be mobile so that it could be moved away from the nodes at these problematic frequencies, see Ref. \cite{geraldes2016new} for example, or multiple sensors would need to be used to reduce the number of cases where there is no detectable motion. In these cases, the voltage produced by the motion of the wire may still be detectable, despite the insensitivity of the displacement sensor.

Another benefit to measuring the emf is that it may be better suited to determine the magnitude of the series coefficients corresponding to higher harmonics of vibration, and therefore to higher driving frequency. Examining the expression for wire displacement, equation \eqref{eq:wireDisp}, it can be seen that $u_n \propto \omega^{-1}$ when $\omega \approx \omega_n$, meaning that the amplitude of the wire vibration will become smaller as the driving frequency increases and, consequently, harder to measure. In comparison, the expression for emf, equation \eqref{eq:emfTime}, predicts the amplitude of the emf to be independent of the driving frequency. This implies that the emf may be better suited for measurements at high driving frequency.

Finally, it may be possible to use this technique for quadrupole fiducialization. By moving the current-carrying wire relative to the quadrupole, the magnetic center may be found by measuring the position where no emf is produced by the wire. A similar approach is described in in Ref. \cite{fischer1992precision}, in which the motion of the wire is mechanically driven. Another attempt at using a vibrating wire for this purpose may be found in Ref. \cite{temnykh1999magnetic}, in which the displacement of the wire is measured instead of the voltage.

In the interest of predicting the sensitivity of the emf to magnetic field strength, one may consider that the derivative of the signal $\fulld \varepsilon / \fulld B_n \propto B_n$, meaning that the sensitivity should vary linearly with the magnitude of the coefficient being measured. A more meaningful approach may be to compare the emf signal to that of the vibration amplitude, the sensitivity of which has already been compared to other magnetic measurement techniques (Ref. \cite{arpaia2015performance}). To predict the relative sensitivity of each technique to the strength of the magnetic field, one may consider the ratio of one measurand to the other. Comparing equations \eqref{eq:ampTem} and \eqref{eq:emfTime}, it can be found that
\begin{equation} \label{eq:sigSensitivity}
\max| \varepsilon_n | \es \dfrac{B_n \omega L}{2}\bsin{\dfrac{n\pi}{L}z_s}^{-1} A_n \ ,
\end{equation}
which, for driving frequencies near a resonant frequency, can be written as
\begin{equation} \label{eq:emfCompAmplitude}
\max| \varepsilon_n | \es \left( n\pi\sqrt{\dfrac{T}{\mu}} \right) \bsin{\dfrac{n\pi}{L}z_s}^{-1} \dfrac{\omega}{\omega_n} \dfrac{B_n}{2} A_n \ \approx \ \left( n\pi\sqrt{\dfrac{T}{\mu}} \right) \bsin{\dfrac{n\pi}{L}z_s}^{-1} \dfrac{B_n}{2} A_n
\end{equation}
upon substitution of equation \eqref{eq:resonantFrequency}. From this it can be seen that when the same magnet is measured using both methods, and the same measurement setup, the emf will be scaled with respect to the vibration amplitude by two factors. The first is a factor proportional to the driving frequency, as has already been noted, and the second is the magnitude of the coefficient being measured, which is the projection of the field onto that particular mode of vibration. This implies that measurement of the emf is less effective for small-magnitude coefficients, unless the frequency of vibration is sufficiently high. In general, equation \eqref{eq:emfCompAmplitude} may be used as a crude way of estimating the response of one approach compared to the other.

\section{Measurement and Analysis} \label{sec:Measurement and Analysis}
\subsection{Measurement}
This section contains a description of the approach used to measure the emf generated by the wire, as well as a discussion of the measurements taken. Due to a shortage of time with the equipment, it was only possible to measure 9 harmonics, and only one measurement per harmonic was possible.

\begin{figure}[H]
\centering
		\includegraphics[width=0.75\textwidth]{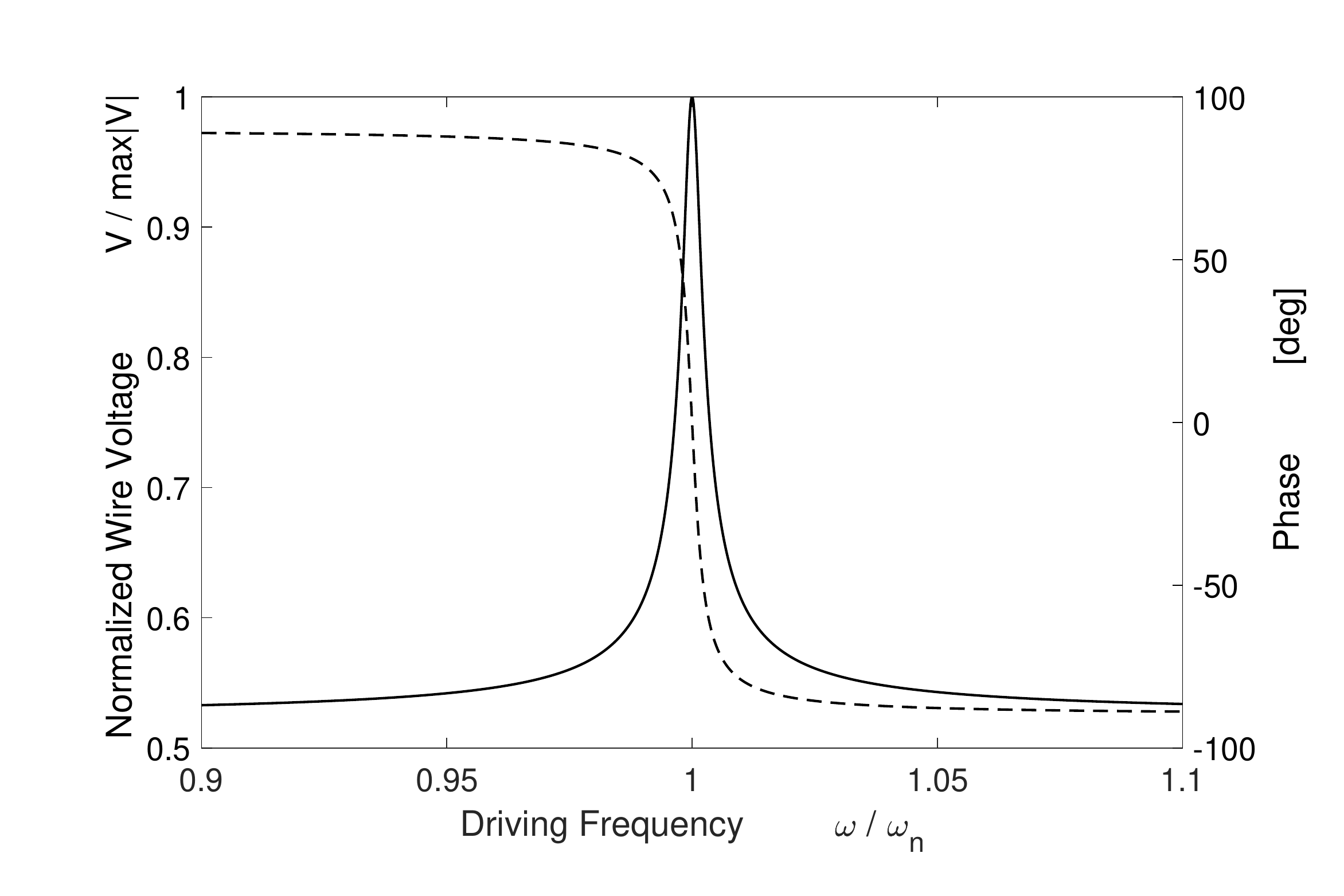}
	\caption{Voltage drop across the wire, predicted using equation \eqref{eq:wireVoltage} with $IR = \SI{1.1}{\volt}$, and the phase shift between the wire and drive voltages, predicted using equation \eqref{eq:emfPhase}.}
	\label{fig:theoreticalCurve}
\end{figure}

In order to test the hypothesis that the emf generated by the vibrating wire may be used to determine the magnetic field acting on the wire, the voltage drop across the wire was measured as a function of driving frequency. This voltage is related to the emf generated by the wire by Kirchoff's Law, 
\begin{equation} \label{eq:wireVoltage}
V \es \varepsilon \ps I R \ ,
\end{equation}
where $V$ is the measured voltage across the wire, $I$ is the drive current, and $R$ is the resistance of the wire. A plot of the predicted form of the measurable quantities, voltage and phase (equations \eqref{eq:wireVoltage} and \eqref{eq:emfPhase}, respectively), can be found in Figure \ref{fig:theoreticalCurve}.

The amplitude of the wire voltage, as well as the phase offset with respect to the driving voltage, was measured using a Signal Recovery model 7265 lock-in amplifier which took the voltage across the wire as a direct input. For these measurements, the magnetic field of an in-vaccum wiggler with a period of \SI{80}{\milli\metre} and a length of about \SI{1.41}{\metre} was used. For the measurement a beryllium-copper wire with a diameter of \SI{0.1}{\milli\metre}, and length of \SI{3.516}{\metre}, was used, and tension was applied using a hanging mass. Parameters of the setup can be found in Table \ref{tab:wireTable}. The data collected may be found in Figure \ref{fig:EMFn}. 

It should be noted that the measurements presented were taken using two different current amplitudes to drive the wire. One set of data, for modes 61 through 70, were collected using a \SI{20}{\milli\ampere} drive current, while a \SI{1}{\milli\ampere} current was used for modes 85 through 90. The choice of current amplitude was made based on the observed amplitude of the vibration near the resonant frequency being considered. The driving force had to be sufficient to produce a clear signal but at the same time not be so large as to excite non-linear phenomena, meaning that frequencies at which the wire was particulary sensitive to driving had to be measured using smaller currents. Those interested in details concerning non-linear phenomena in the motion of vibrating wires may refer to \cite{narasimha1968non, hanson1994measurements}, and especially \cite{pedersen2017direct}.

\begin{table}[H]
\centering
\begin{tabular}{| c | c | l |}
\hline
Parameter & Symbol & Value \\
\hline
Density & $\mu$ & \SI{71.74}{\milli\gram \cdot \metre^{-1}} \\
Length & $L$ & \SI{3.516}{\metre}  \\
Tension & $T$ & \SI{9.022}{\newton} \\
Fundamental Frequency & $f_0$ & \SI{51.6}{\hertz} \\
\hline
\end{tabular}
\caption{Parameters of the experimental setup.}
\label{tab:wireTable}
\end{table}

\subsection{Analysis}
\begin{figure}[H]
\centering
	\begin{subfigure}{0.495\textwidth}
		\includegraphics[width=1.0\textwidth]{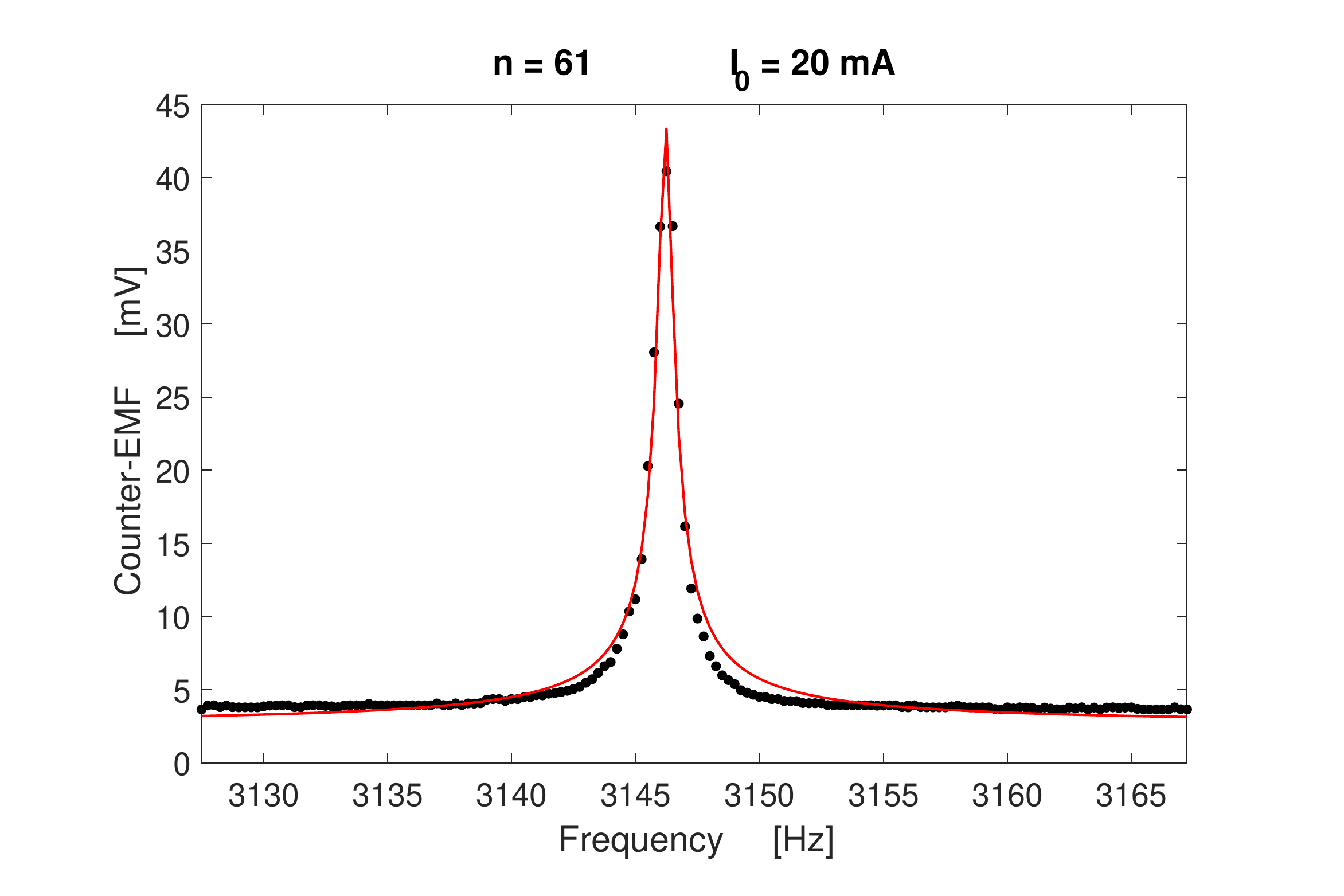}
	\end{subfigure}
	\begin{subfigure}{0.495\textwidth}
		\includegraphics[width=1.0\textwidth]{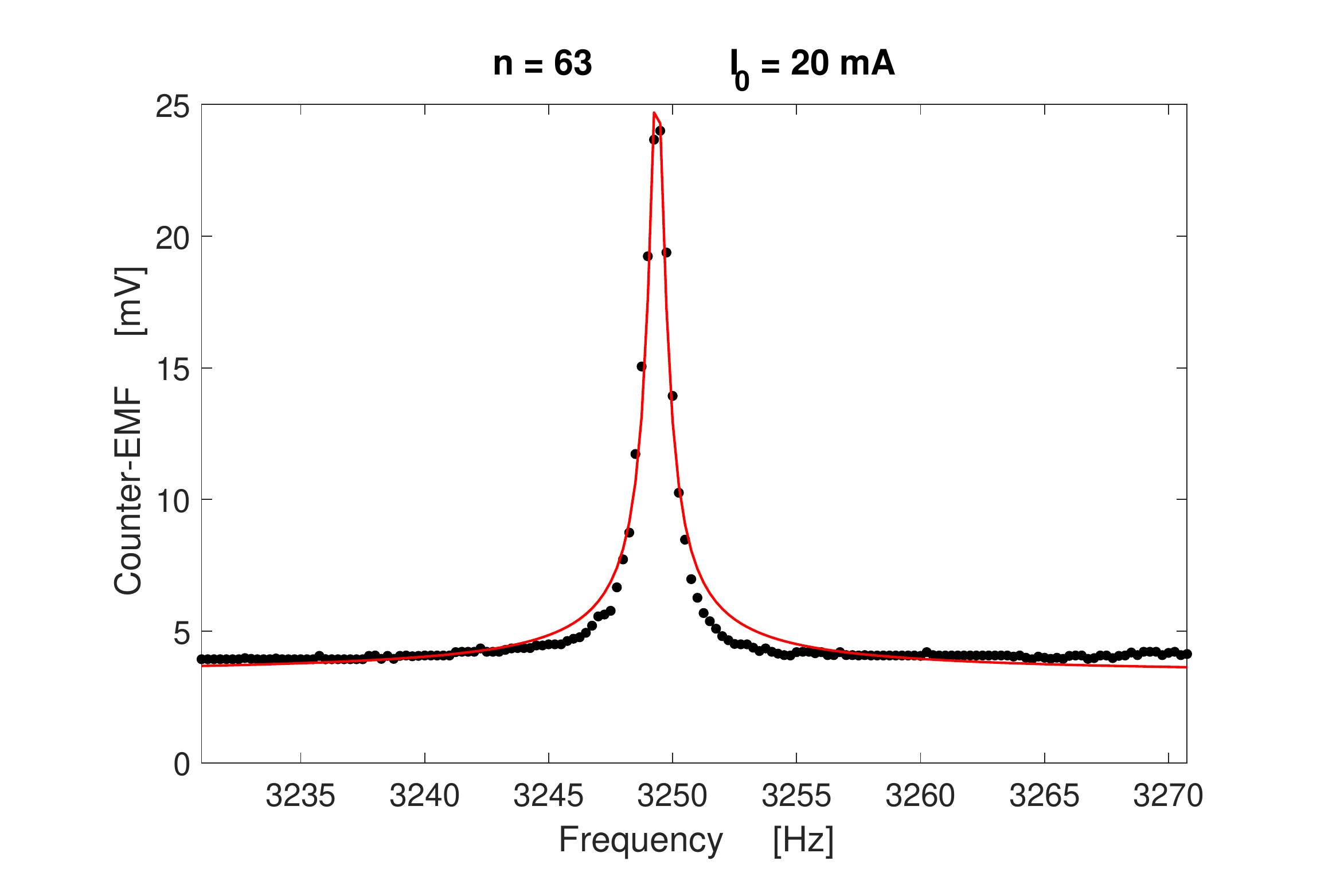}
	\end{subfigure}
	
	\begin{subfigure}{0.495\textwidth}
		\includegraphics[width=1.0\textwidth]{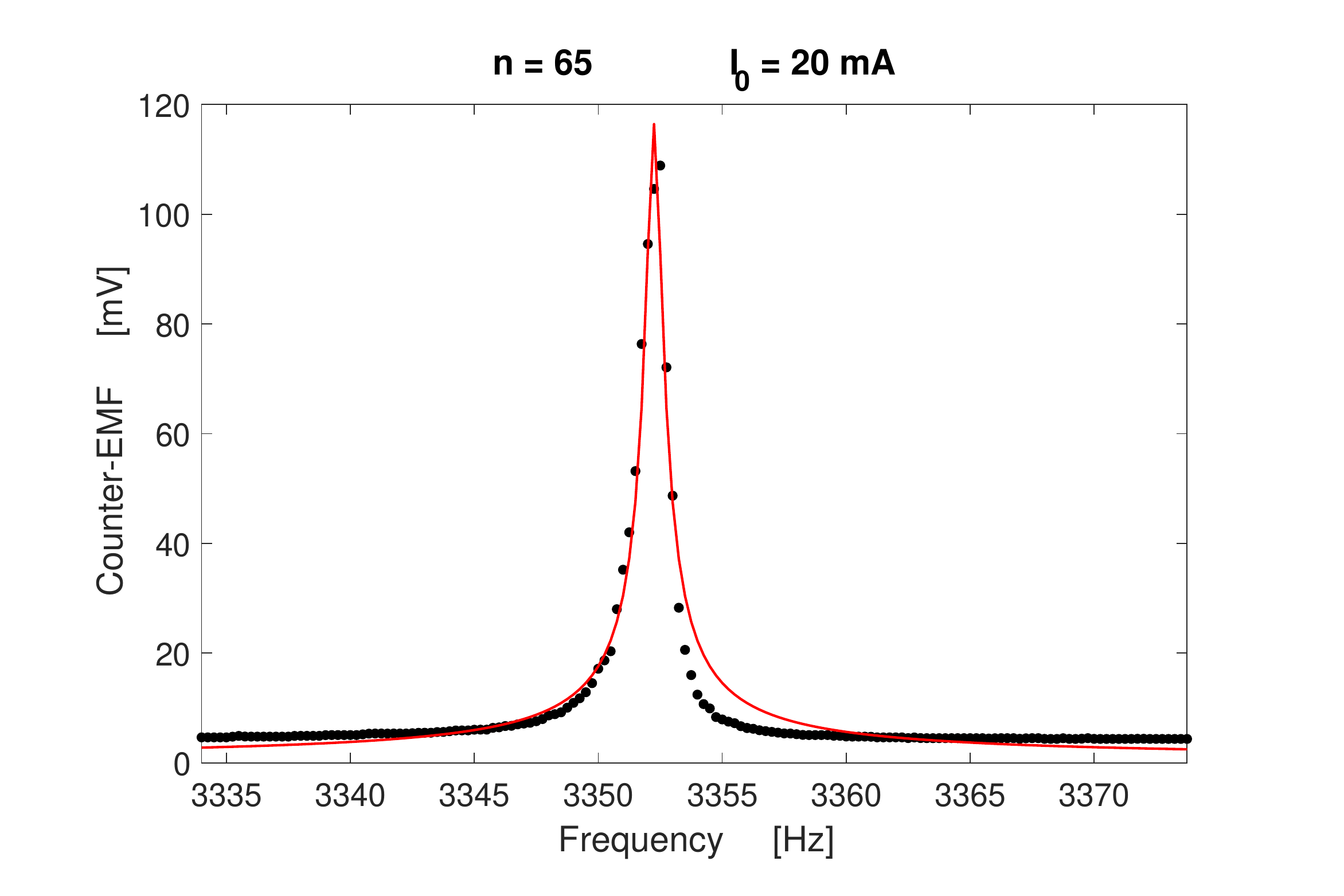}
	\end{subfigure}
	\begin{subfigure}{0.495\textwidth}
		\includegraphics[width=1.0\textwidth]{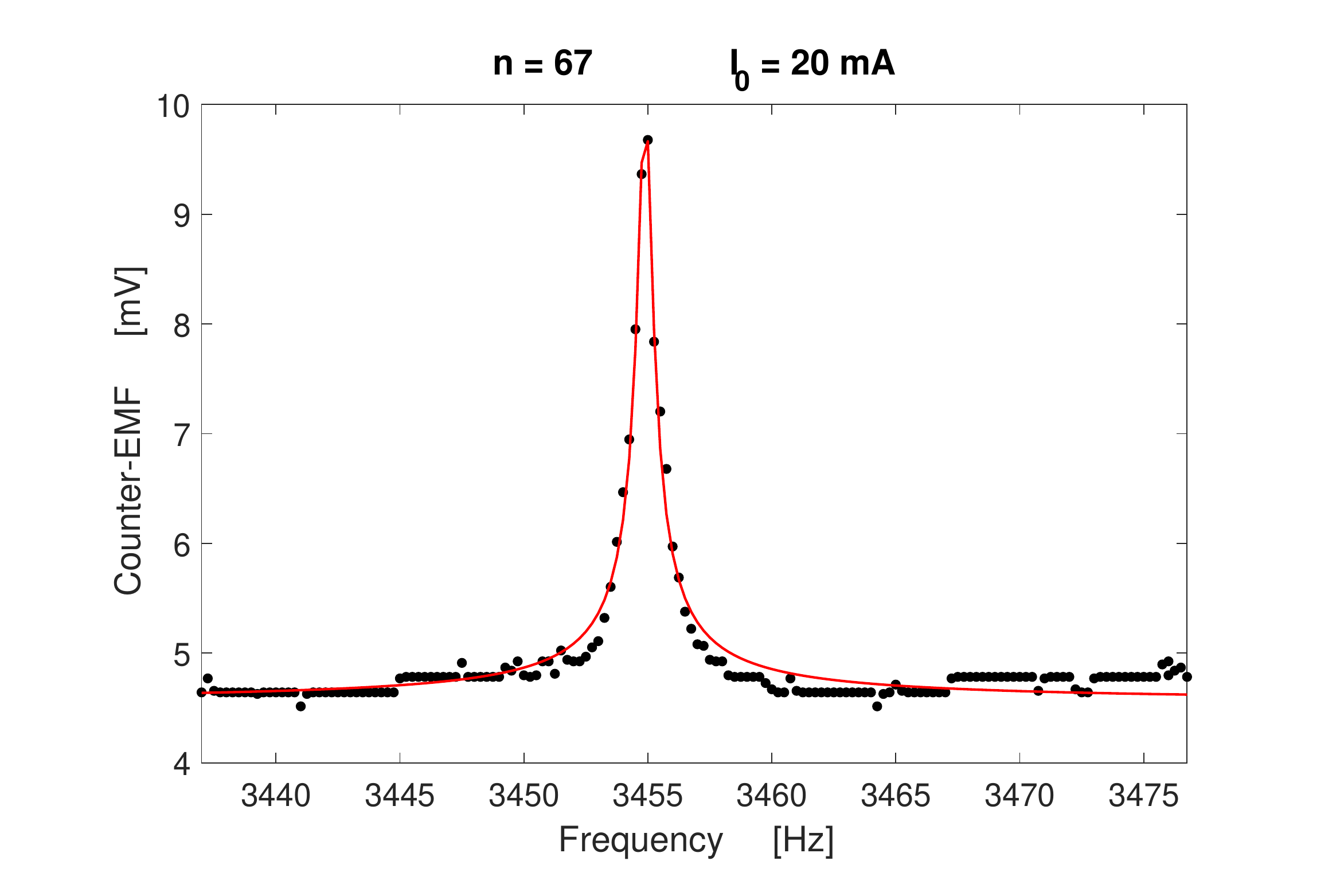}
	\end{subfigure}
	
	\begin{subfigure}{0.495\textwidth}
		\includegraphics[width=1.0\textwidth]{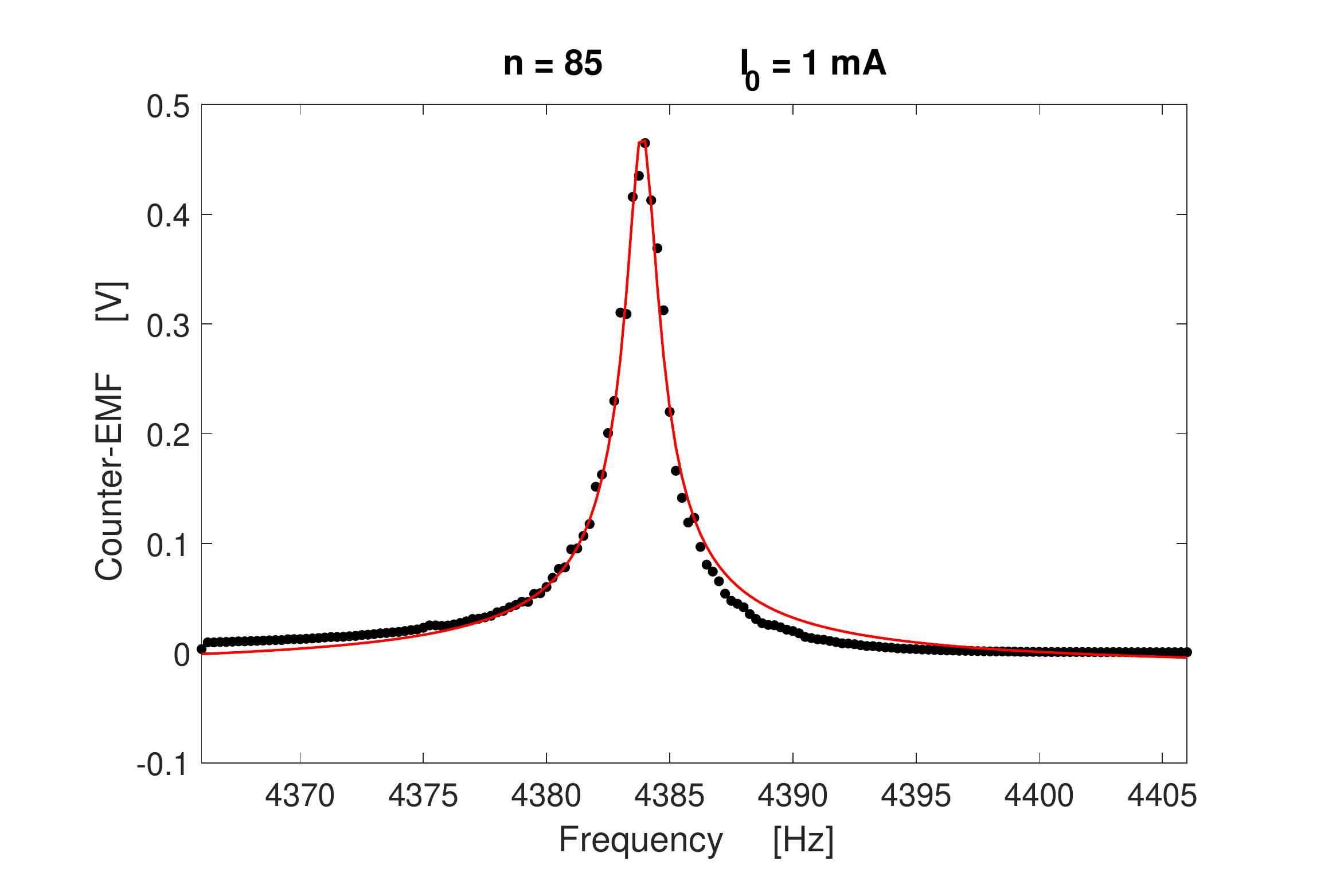}
	\end{subfigure}
	\begin{subfigure}{0.495\textwidth}
		\includegraphics[width=1.0\textwidth]{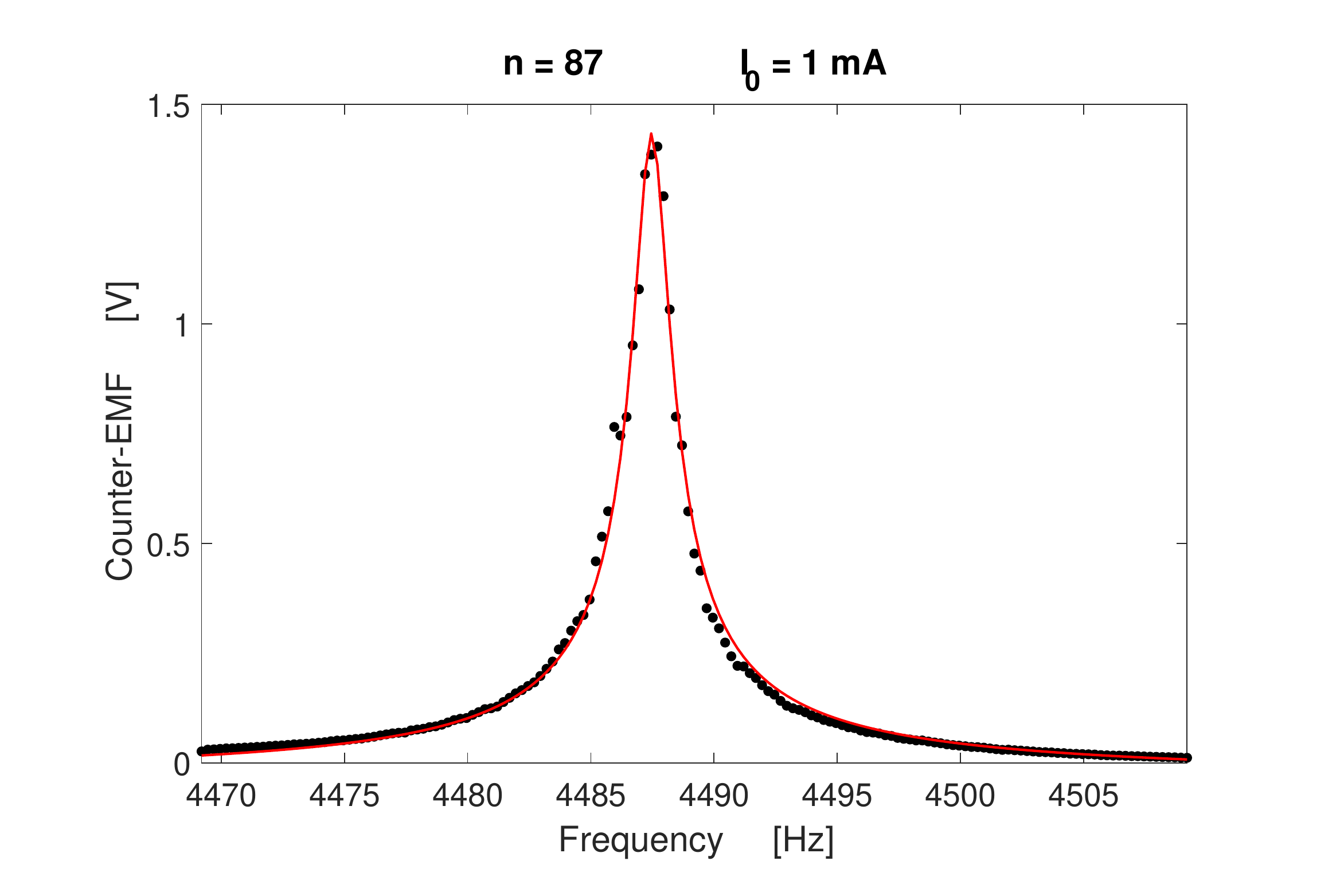}
	\end{subfigure}
	
	\begin{subfigure}{0.495\textwidth}
		\includegraphics[width=1.0\textwidth]{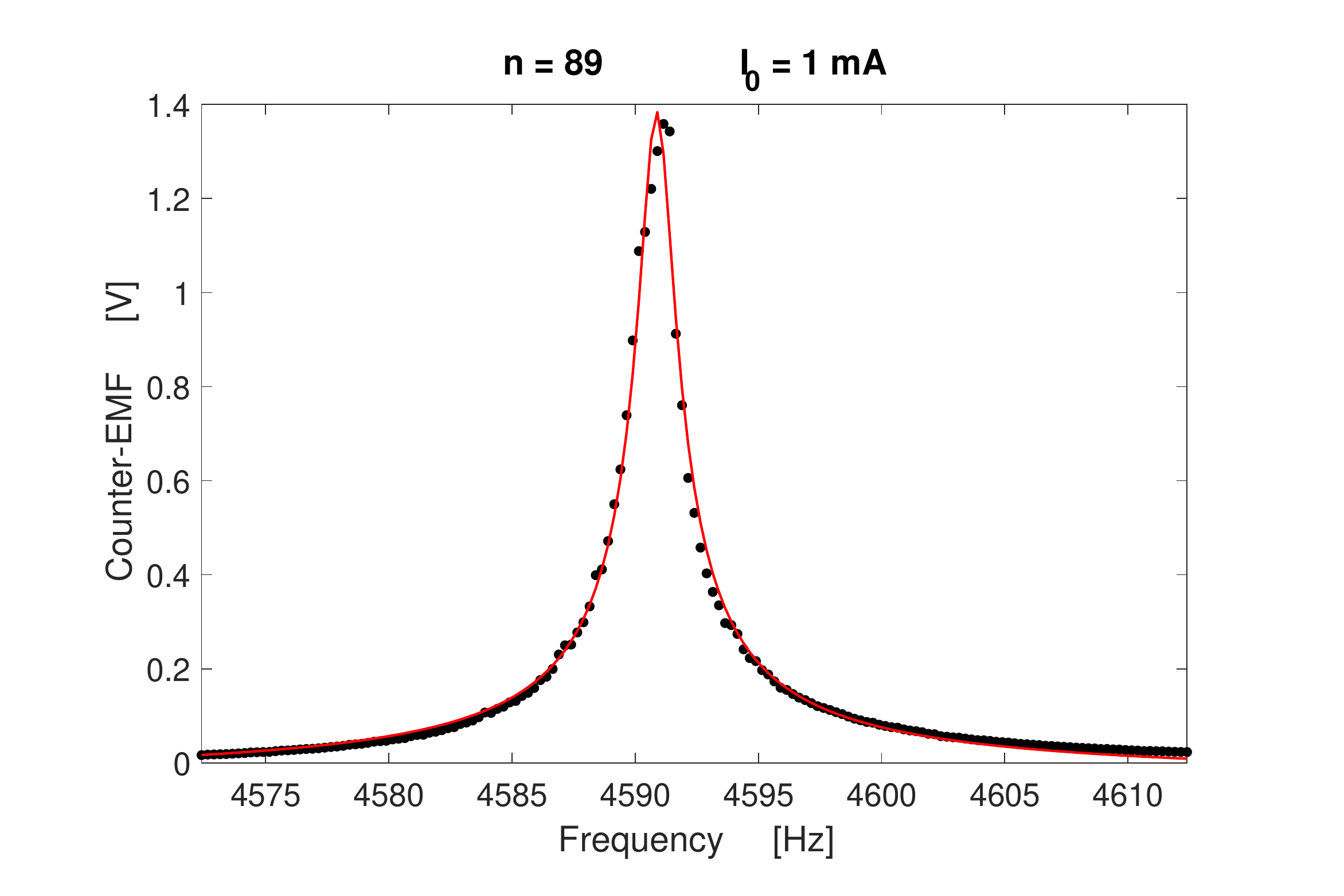}
	\end{subfigure}
	\begin{subfigure}{0.495\textwidth}
		\includegraphics[width=1.0\textwidth]{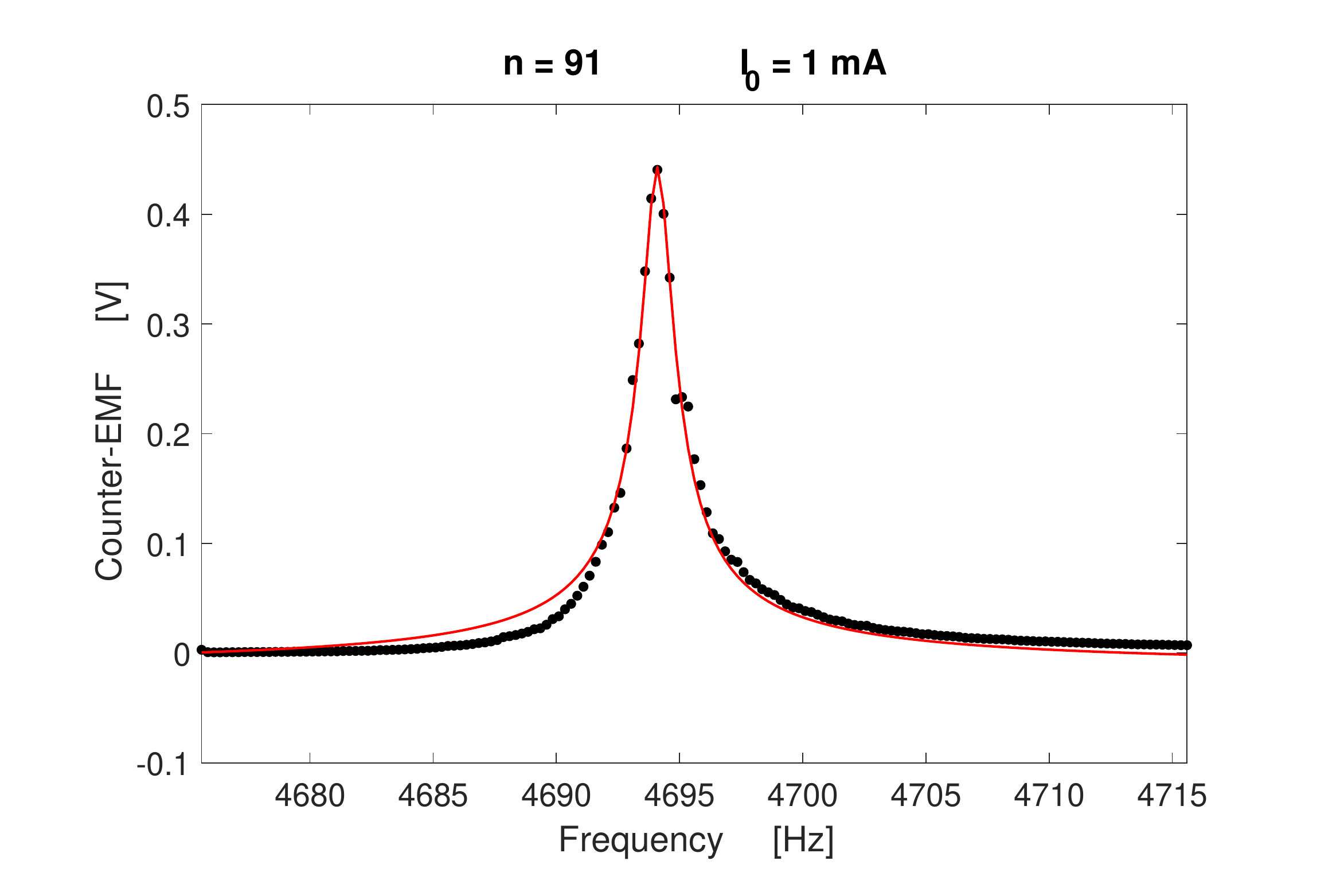}
	\end{subfigure}
	
\caption{Measurements of emf with theoretical curve, equation \eqref{eq:emfFitOff}, fit to data. $n$ is the harmonic number, and $I_0$ is the amplitude of the driving current.}
\label{fig:EMFfit}
\end{figure}

The measurements of the wire voltage were then compared to the theory outlined in Section \ref{sec:EMF} by fitting the following altered version of equation \eqref{eq:emfFit}. In order to account for variations in the driving voltage of the wire, such as those caused by a change in ambient temperature, a DC offset parameter was introduced to the curve-fitting equation so that
\begin{equation} \label{eq:emfFitOff}
\varepsilon_n(\omega) \es a_n \omega\left[ c_n\omega^2 - (\omega^2 - b_n )^2 \right]^{-1/2} \ps d_n \ .
\end{equation} 
The result of this curve fitting process can be found in Figure \ref{fig:EMFfit}. A measure of the quality of the fit for each data set can be found in Table \ref{tab:errTable}. The quantity $R$ found there is the sum of square differences between the data set and the fit, normalized by the sum of squared measurements. This normalization is applied to ensure the measure is independent of the scale of the data. Expressed mathematically,
\begin{equation} \label{eq:fitMeasure}
R \es \left( \sum_k {y_k}^2 \right)^{-1} \sum_k \left[ y_k - f(\omega_k) \right]^{2} \ ,
\end{equation} 
where $y_k$ is the $k^{th}$ element of the data set and $f(\omega_k)$ is the fit-function evaluated at the $k^{th}$ frequency measured.

\begin{figure}[H]
\centering
	\begin{subfigure}{0.495\textwidth}
		\includegraphics[width=1.0\textwidth]{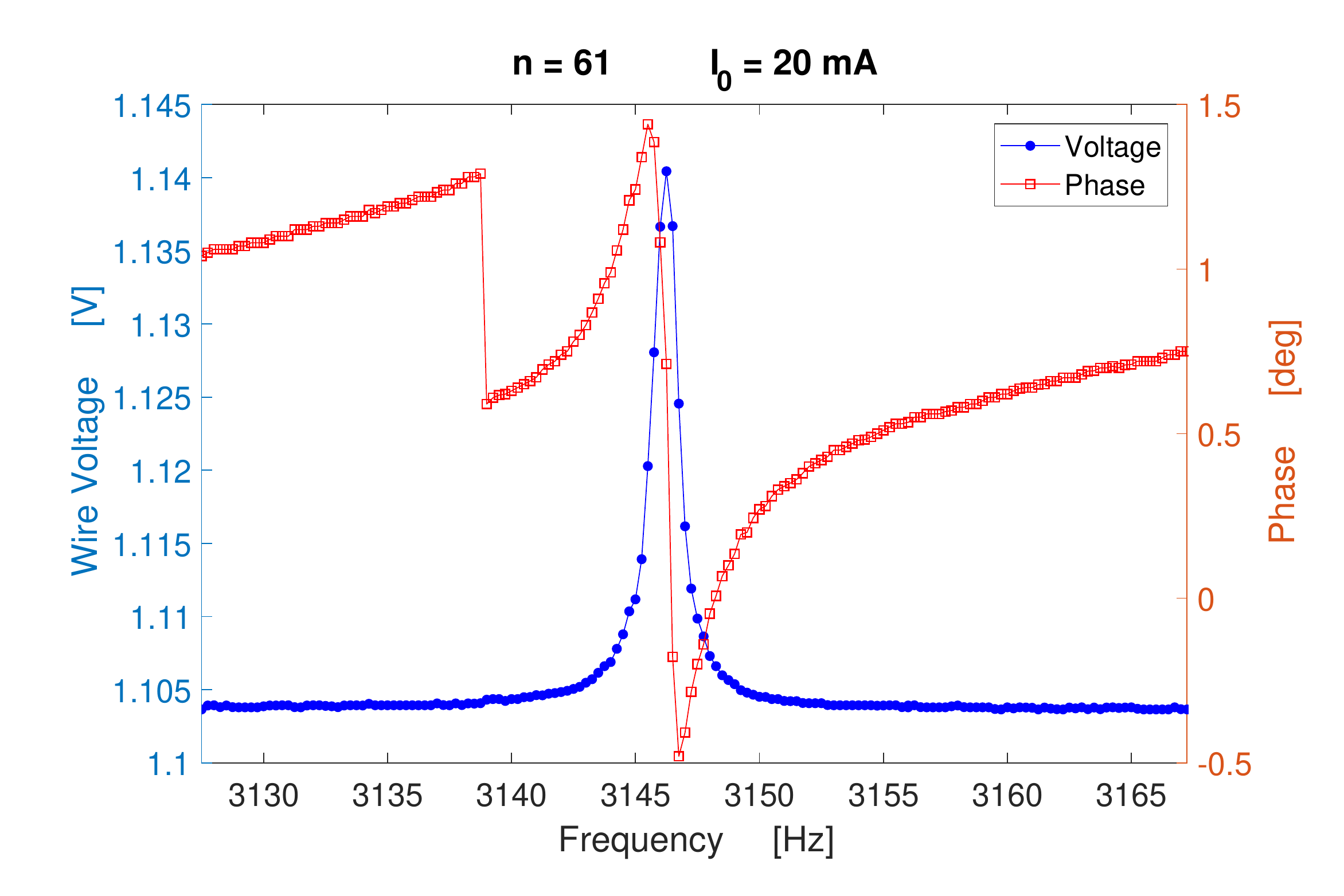}
	\end{subfigure}
	\begin{subfigure}{0.495\textwidth}
		\includegraphics[width=1.0\textwidth]{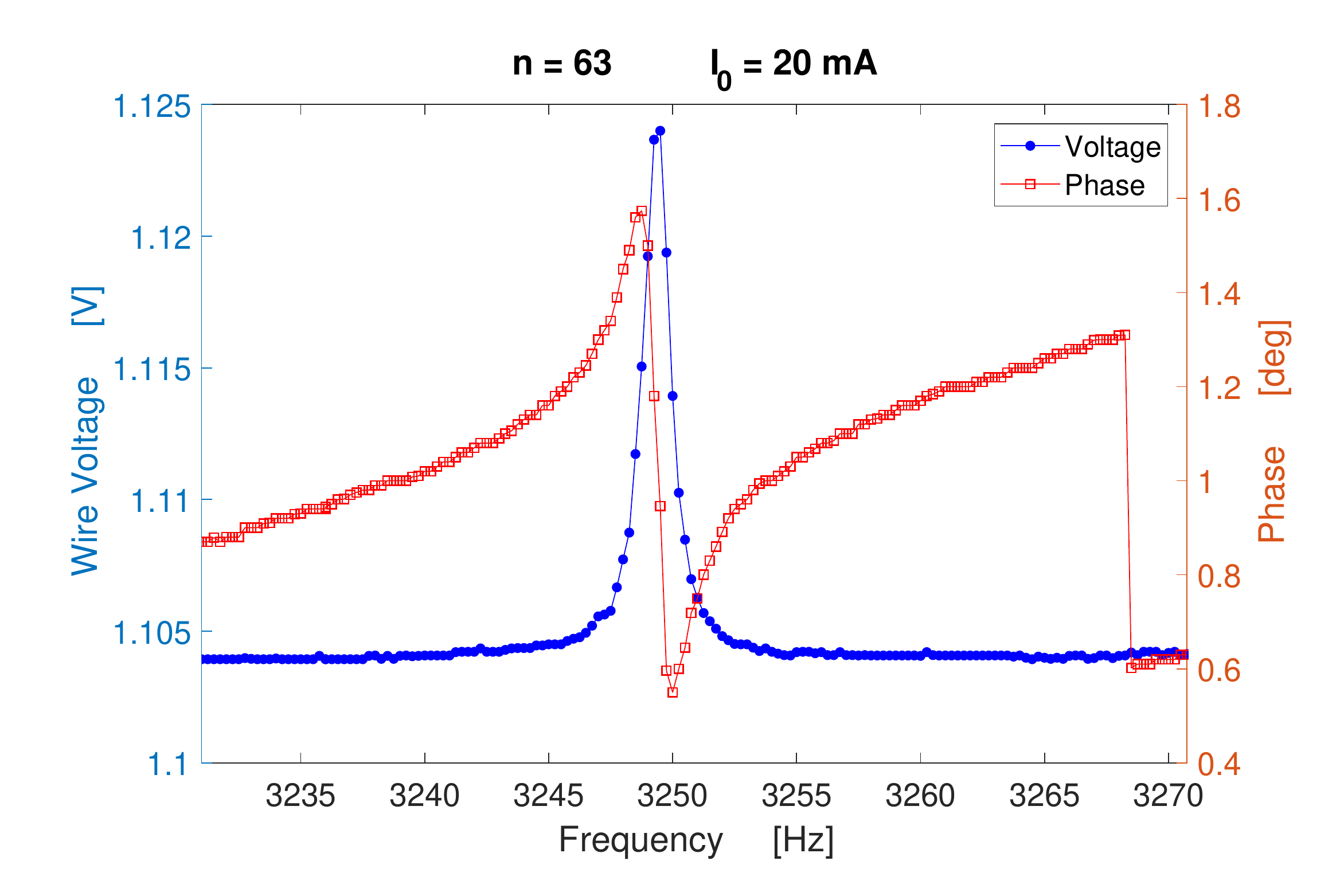}
	\end{subfigure}
	
	\begin{subfigure}{0.495\textwidth}
		\includegraphics[width=1.0\textwidth]{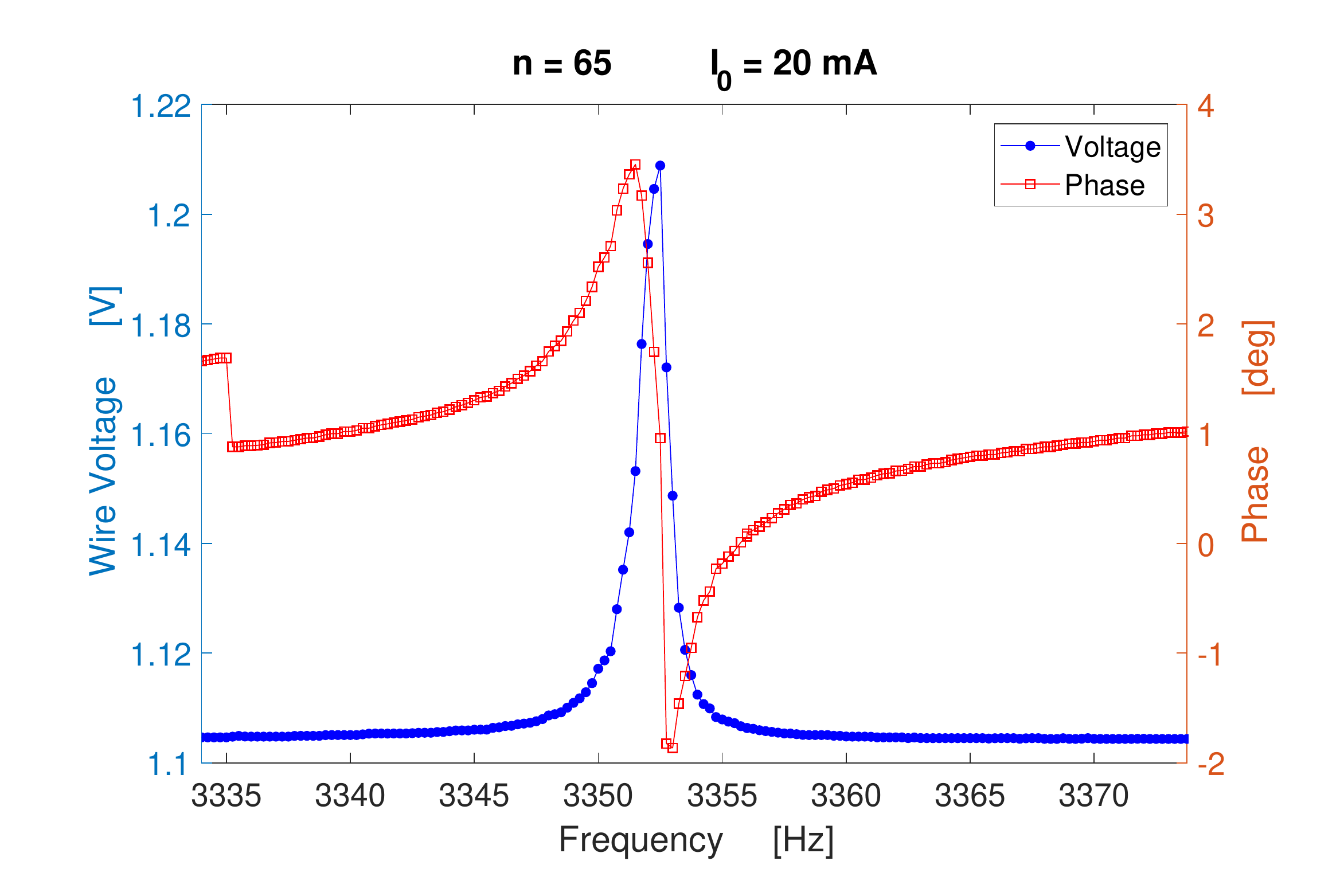}
	\end{subfigure}
	\begin{subfigure}{0.495\textwidth}
		\includegraphics[width=1.0\textwidth]{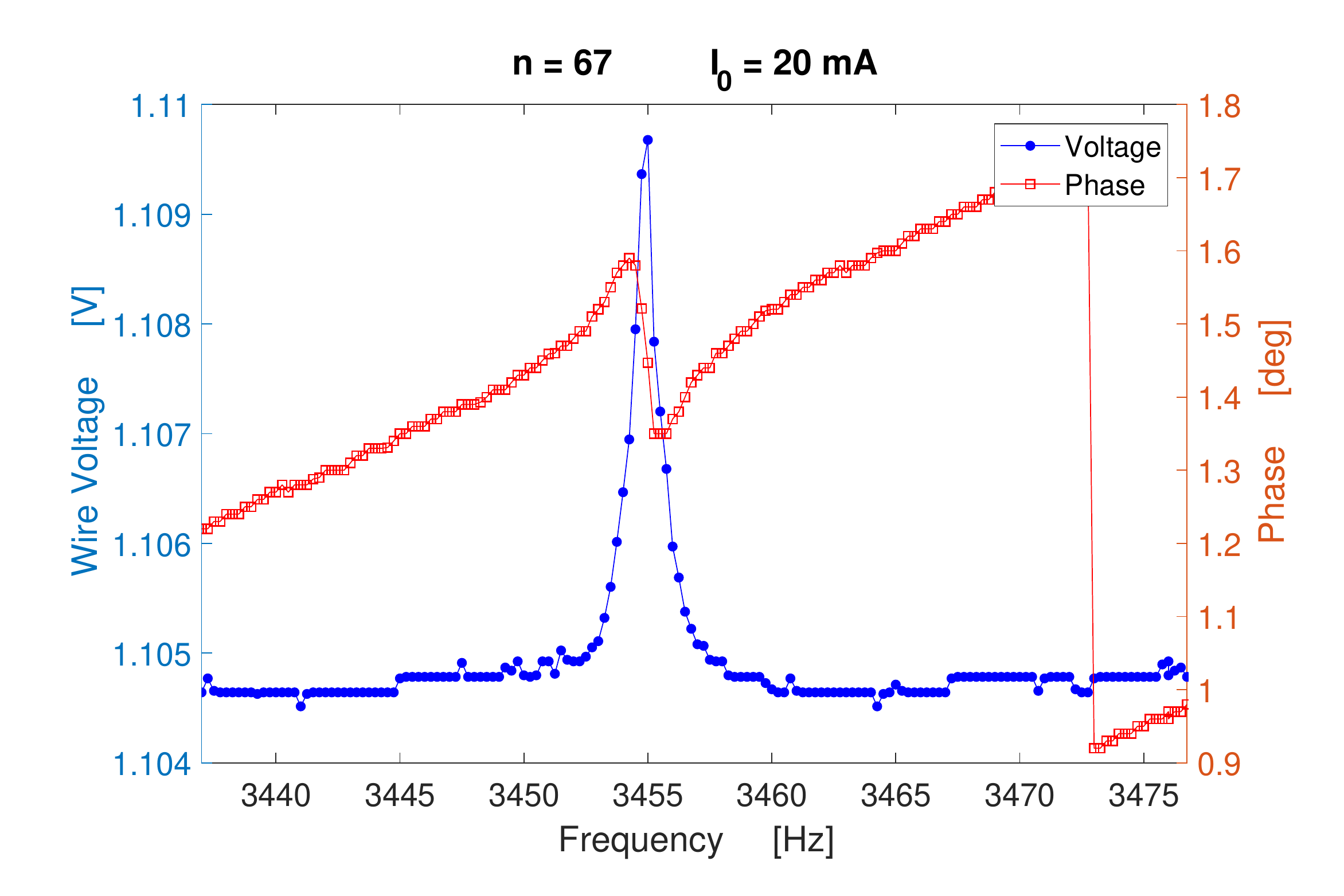}
	\end{subfigure}
	
	\begin{subfigure}{0.495\textwidth}
		\includegraphics[width=1.0\textwidth]{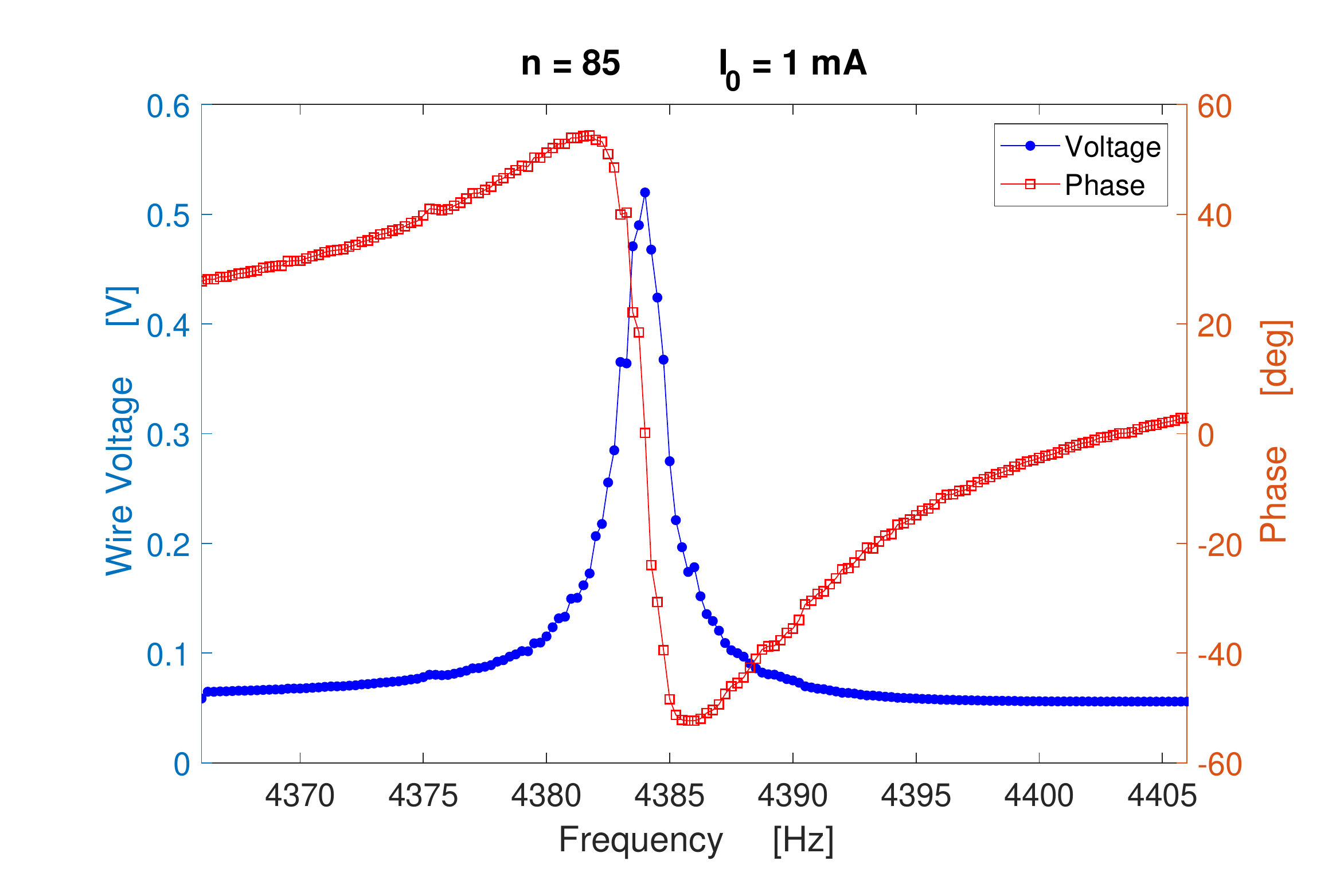}
	\end{subfigure}
	\begin{subfigure}{0.495\textwidth}
		\includegraphics[width=1.0\textwidth]{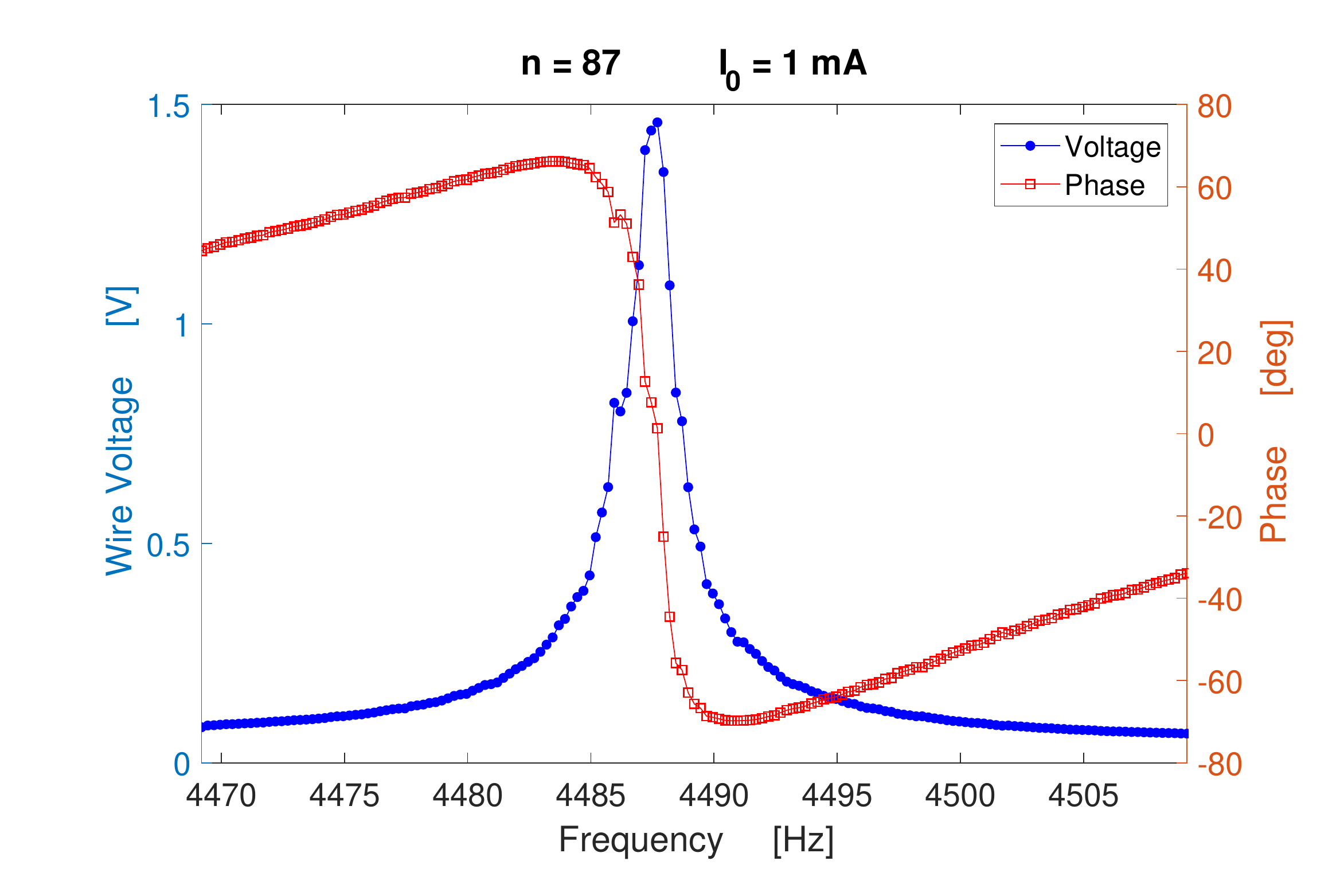}
	\end{subfigure}
	
	\begin{subfigure}{0.495\textwidth}
		\includegraphics[width=1.0\textwidth]{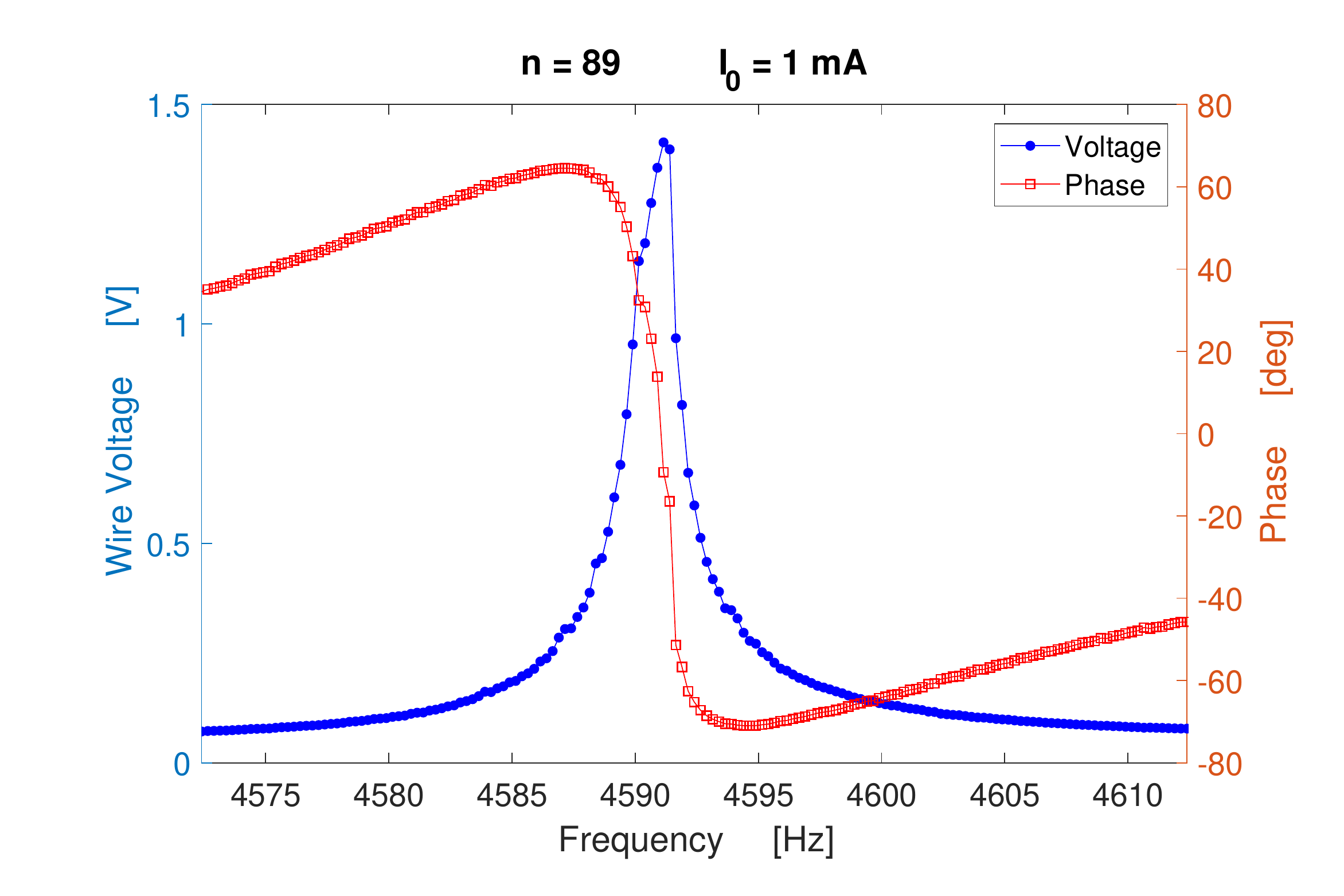}
	\end{subfigure}
	\begin{subfigure}{0.495\textwidth}
		\includegraphics[width=1.0\textwidth]{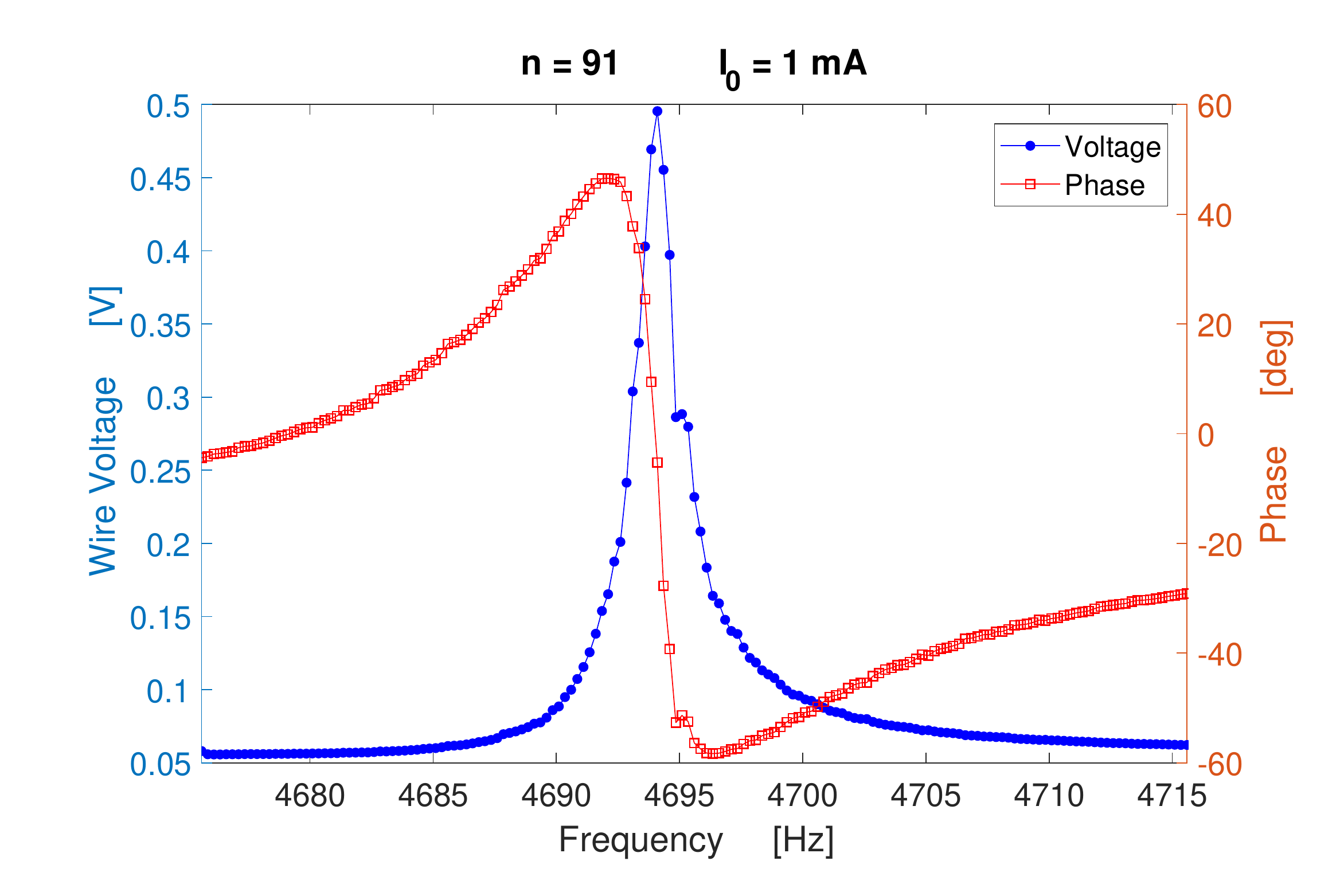}
	\end{subfigure}
	
\caption{Measurements of wire voltage for several resonant frequencies. $n$ is the harmonic number, and $I_0$ is the amplitude of the driving current.}
\label{fig:EMFn}
\end{figure}

Examining the data illustrated in Figure \ref{fig:EMFn}, it can be seen that, in general, the measured curves are of the same shape as the predicted curves found in Figure \ref{fig:theoreticalCurve}. However, there are some features which appear unique to the driving current used. In particular, the measurements taken using a drive current of \SI{20}{\milli\ampere} appear to have an irregular phase shift, in that the phase is bound between $\pm\SI{10}{\degree}$ while the measurements taken using a drive current of \SI{1}{\milli\ampere} have a phase bound between $\pm\SI{60}{\degree}$. Referring to Figure \ref{fig:theoreticalCurve}, it appears the latter case agrees more closely with theory. This implies either an unaccounted for dependence on the drive current, or an experimental error. It seems likely that this difference is due to measurement error related to the difference in signal strength. Typically, the curves measured at a smaller drive current have heights between 0.5 and 1 volts, while the other curves are an order of magnitude smaller, and the latter are notably narrower than the former. If measurement error is responsible for this discrepancy, it could possibly be rectified by sampling the curve at a higher frequency resolution.

To check the accuracy of the series coefficients calculated using measurements of the emf $B_{n\varepsilon}$, the values of which may be found in Table \ref{tab:errTable}, comparison was made to those calculated from measurements of wire displacement $B_{nA}$. In theory, the two approaches to measurement should give the same series coefficients. It should be noted that this comparison makes use of data sets collected several weeks apart. In this time the apparatus was used intermittently, and the same wire remained strung. Based on observations made throughout the use of the vibrating wire, it does not seem likely that the time interval between measurements will affect the values of the series coefficients.

\begin{figure}[H]
\centering
		\includegraphics[width=0.75\textwidth]{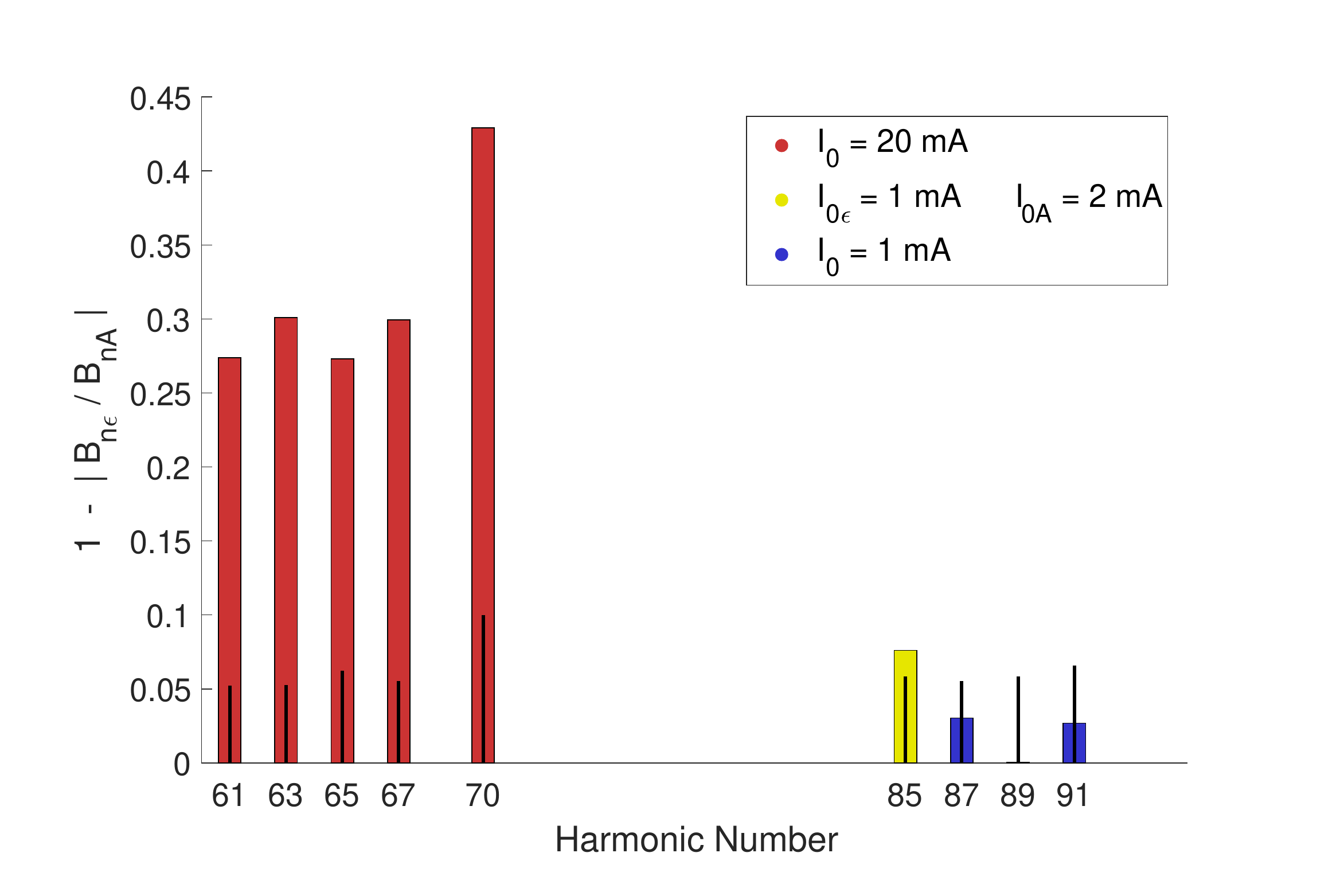}
	\caption{Plot of relative error in emf coefficients. Bars coloured to indicate data set. $I_{0\epsilon}$ indicates driving current used for voltage measurements, while $I_{0A}$ indicates that used for displacement measurements. For those labelled by $I_0$, $I_{0\epsilon} = I_{0A}$. Black bars represent the uncertainty in the value. (If viewed in grayscale, Table \ref{tab:errTable} on p.\pageref{tab:errTable} may be used as reference).}
	\label{fig:coeffErr}
\end{figure}

A comparison of the two sets of coefficients may be found in Table \ref{tab:errTable}, and is visualized in  Figure \ref{fig:coeffErr}, which shows the relative difference between the two sets of coefficients, taking $B_{nA}$ as the reference. These figures show close agreement of coefficients 87, 89, and 91, as well as the apparent relationship between driving current and measurement discrepancy. To elaborate on the latter claim, comparison of the two sets of coefficients showed that $B_{n\varepsilon}$ measured at \SI{20}{\milli\ampere} differed from $B_{nA}$. It is noteworthy that these $B_{n\varepsilon}$ correspond to data sets with irregular phase shifts, seen in Figure \ref{fig:EMFn}. In addition, the $B_{n\varepsilon}$ for $n=85$ is exceptional in that it was measured using a different drive current than the $B_{nA}$ it is being compared to. If there is an unaccounted for dependence on current, this comparison may be invalid. 

Examining Figure \ref{fig:EMFn}, it is interesting to note that in the cases of $n=89$ and $n=91$ some signs of non-linear oscillation, resulting from excessive vibration amplitude, may be seen. The main feature which implies non-linear oscillation is the asymmetry of the amplitude curves, namely that there appears to be a sharp drop on the high-frequency side of the curve associated with the collapse of transverse whirling motion \cite{hanson1994measurements}. Even though the motion of the wire appears to be outside the regime described by equation \eqref{eq:diffEq}, the series coefficients calculated using either method are in close agreement. This would indicate that non-linear wire motion may have either a similar, or minor, impact on the values of the series coefficients. In the least, this would mean that the measurement of emf is no more sensitive to this problematic oscillation than the measurement of displacement.

\begin{table}[H]
\centering
\begin{tabular}{c c c rcl c c c }
\hline
$n$ & $I_{0\varepsilon} (\si{\milli\ampere})$ & $I_{0A} (\si{\milli\ampere})$ & \multicolumn{3}{c}{$B_{n\varepsilon} (\si{\gauss})$} & $B_{nA} (\si{\gauss})$ & $\delta B_n (\%)$ & $R (10^{-4})$ \\
\hline
61 & 20 & 20 & 175.1 &$\pm$& 9.3 & 241.2 & 27 & 12  \\
63 & 20 & 20 & 130.3 &$\pm$& 7.3 & 186.3 & 30 & 7.1 \\
65 & 20 & 20 & 312 &$\pm$& 21 & -429.0 & 27 & 28 \\
67 & 20 & 20 & 62.7 &$\pm$& 3.8 & 89.42 & 30 & 0.63 \\
70 & 20 & 20 & 49.5 &$\pm$& 7.7 & 86.67 & 43 & 1.80 \\
85 &  1  & 2  & 39590 &$\pm$& 170 & -42850 & 8 & 11 \\
87 &  1  & 1  & 73810 &$\pm$& 260 & 71640 & 3 & 5.5 \\
89 &  1  & 1  & 73310 &$\pm$& 290 & -73280 & 0 & 7.3 \\
91 &  1  & 1  & 37550 &$\pm$& 180 & 38590 & 3 & 13 \\
\hline
\end{tabular}
\caption{Table of drive currents used for each measurement set, coefficients calculated for each set, and corresponding percent difference between $B_{n\varepsilon}$ and $B_{nA}$. $B_{nA}$ are reported to arbitrary precision. $R$ is a measure of the quality of the curve fit, normalized to be independent of the scale of the data.}
\label{tab:errTable}
\end{table}

\section{Conclusion} \label{sec:Conclusion}

In Section \ref{sec:EMF} a theory of an alternative way of measuring magnetic fields using the vibrating wire technique was developed. This alternative makes use of the electromotive force generated by the motion of the wire in the field. The advantages, disadvantages, and potential applications of this theory were discussed in Section \ref{sec:Application}. In Section \ref{sec:Measurement and Analysis} experimental tests of the theory are reported, the results of which were found to agree with the conventional method for 3 of 9 test cases. The cases which did not agree used a different drive current than the others, and produced considerably smaller signals with narrower resonance curves. Based on this it is thought that the discrepancy between the remaining 6 cases may be due to a much smaller measured signal, the amplitude of the driving current used, or to the frequency resolution at which the curves were measured. 

For comparison, the advantages and disadvantages of the technique with respect to the conventional method will be summarized. Among the advantages is the lack of need for a displacement sensor, which means simpler experimental design and setup. Another is, theoretically, a stronger signal at high frequencies of vibration. The disadvantages include a lack of information on the sign of the series coefficient, making the method unsuitable for determining field distributions, as well as the fact that the orientation of the field under measurement is ambiguous. In addition, the sensitivity of the technique is predicted to be more complicated than the conventional technique, in that, comparatively, the sensitivity of the emf is worse for small-magnitude coefficients, but becomes better as the magnitude increases.

To conclude, based on the results presented here, it would appear that the theory developed in Section \ref{sec:EMF} is  valid. This means that, in principle, it may be possible to determine magnetic field through measurements of the voltage across the vibrating wire instead of the amplitude of the wire vibration. A more thorough analysis will be needed to assess the practicality of this technique, as well as identify unaccounted for factors, such as non-linear current dependence.

\section{Acknowledgements}
The authors are grateful to the following people. Jon Stampe for valuable discussion and aid in experimental setup. Tor Pederson, Grant Henneberg, Garth Steel, Bruce Wu and Carl Jansen for aid in developing the measurement system. 

Research at the Canadian Light Source was funded by the Canada Foundation for Innovation, the Natural Sciences and Engineering Research Council of Canada, the National Research Council Canada, the Canadian Institutes of Health Research, the Government of Saskatchewan, Western Economic Diversification Canada, and the University of Saskatchewan.
 
\section{Author Contributions}
JG planned and carried out measurements, developed the model, analysed the data, and wrote the manuscript. CB contributed to the analysis of the data used as reference, and to the editing of the final manuscript.

\newpage
\section{Figures}

\newpage

\newpage

\newpage
\bibliographystyle{unsrt}
\bibliography{BibList}

\end{document}